\documentstyle[12pt]{article}

\newcommand {\e} {\mbox{\rm e}}

\newcommand {\nn}    {\nonumber}
\newcommand {\vs}[1]  { \vspace*{#1 cm} }

\newcounter{eq}
\newcounter{sc}

\newcommand {\MPL}  {Mod. Phys. Lett.}
\newcommand {\IJMP}  {Int. J. Mod. Phys.}

\newcommand {\NPPS}  {Nucl. Phys. Proc. Suppl.}
\newcommand {\PL}   {Phys. Lett.}
\newcommand {\PR}   {Phys. Rev.}
\newcommand {\PRL}   {Phys. Rev. Lett.}
\newcommand {\CMP}  {Comm. Math. Phys.}
\newcommand {\AP}   {Ann. of Phys.}
\newcommand {\PTP}  {Prog. Theor. Phys.}

\newcommand {\CQG}  {Class. Quantum. Grav.}

\def\overleftrightarrow#1{\vbox{\ialign{##\crcr
 $\leftrightarrow$\crcr\noalign{\kern-1pt\nointerlineskip}
 $\hfil\displaystyle{#1}\hfil$\crcr}}}

\newlength{\minitwocolumn}
\begin{document}


\begin{flushright}
EDO-EP-47\\
November, 2003\\
\end{flushright}
\vspace{30pt}

\pagestyle{empty}
\baselineskip15pt

\begin{center}
{\large\bf A Relation Between Topological Quantum Field Theory 
\vskip 1mm and the Kodama State 
 \vskip 1mm
}

\vspace{20mm}

Ichiro Oda
          \footnote{
          E-mail address:\ ioda@edogawa-u.ac.jp
                  }
\\
\vspace{10mm}
          Edogawa University,
          474 Komaki, Nagareyama City, Chiba 270-0198, JAPAN \\

\end{center}


\vspace{15mm}
\begin{abstract}
We study a relation between topological quantum field theory and the 
Kodama (Chern-Simons) state. It is shown that the Kodama (Chern-Simons) 
state describes a topological state with unbroken diffeomorphism invariance
in Yang-Mills theory and Einstein's general relativity in four dimensions.
We give a clear explanation of "why" such a topological state exists.

\vspace{15mm}

\end{abstract}

\newpage
\pagestyle{plain}
\pagenumbering{arabic}


\rm
\section{Introduction}

It has been known for a long time that Yang-Mills theory in four
dimensions has an exact zero energy state of the Schrodinger
equation \cite{Jackiw}, which is, what we call, the Chern-Simons
state, and is expressed by the exponential of the Chern-Simons form
\begin{eqnarray}
\Psi = \exp( \pm c S_{CS}),
\label{1.1}
\end{eqnarray}
with $c$ being a suitable constant and $S_{CS}$ being explicitly given by
$S_{CS} = \int Tr ( \frac{1}{2} A \wedge dA + \frac{1}{3} A \wedge A \wedge A )$. 
On the other hand, in the community of loop quantum gravity,
this state is called the Kodama state since Kodama has first pointed out 
that the exponential of the Chern-Simons form solves the quantum Ashtekar 
constraints \cite{Ashtekar} by starting from the solution
for Bianchi IX model and generalizing it \cite{Kodama}. This state
has been extensively investigated by Smolin \cite{Smolin} since it
shows that, at least for de Sitter space-time, loop quantum gravity
{\it{does}} have a good low energy limit, thereby reproducing familiar
general relativity and quantum field theory at the low energy as desired.

Recently, Witten has emphasized that the Chern-Simons state of Yang-Mills
theory could not be the ground state of the theory since this state
is highly unnormalizable and is not invariant under CPT, and in addition
negative helicity states have not only negative energy but also negative
norm in expanding around this state \cite{Witten1}. In response to Witten's 
paper from the loop gravity side, Freidel and Smolin have argued that the Kodama
state is delta-functional normalizable in the Euclidean theory while
it is not normalizable in the Lorentzian theory with the kinematical
inner product \cite{Freidel}. They have also discussed that there is still 
the logical possibility such that the Kodama state in the Lorentzian gravity 
theory might become normalizable if we take into account the contribution to
the physical inner product from the measure to all order. More recently,
a relation of self-duality and the Kodama state is studied in the
abelian gauge theory \cite{Corichi}.

The purpose of this article is to point out an interesting relationship between
topological quantum field theory and the Kodama (Chern-Simons) state. 
To be precise, we will show the statement that the Kodama (Chern-Simons) state 
exists whenever a theory can be rewritten to the second Chern class
$S = \int Tr F \wedge F$ under appropriate ansatze. In this context,
we can clearly understand "why" such a topological state exists in Yang-Mills
theory and general relativity in four dimensions which was one of
motivations in the Witten's paper \cite{Witten1}. 

This paper is organized as follows. In Sec. 2 we review Horowitz's
work \cite{Horowitz} where it was mentioned that the Kodama (Chern-Simons) state 
also exists in topological quantum field theory in four dimensions 
\footnote{In this sense, it might be more suitable to call the state 
under consideration "Kodama-Chern-Simons-Horowitz state".}. In Sec. 3,
we show that the Kodama (Chern-Simons) state exists if we can cast a theory into 
the form of the second Chern class $S = \int Tr F \wedge F$. This fact
implies that the Kodama (Chern-Simons) state is a topological state
which describes a unbroken phase of diffeomorphism invariance. We conclude
in Sec. 4 with a discussion of the results obtained in this article.

\section{ Review of Horowitz's topological quantum field theory}

We begin with one of the simplest examples where there is a unique quantum state 
(up to overall constant).  This model has been already constructed by 
Horowitz \cite{Horowitz} and clearly presents the reason why a unique 
quantum state exists in topological quantum field theory of cohomological 
type \cite{Witten2}.

Let us consider an arbitrary scalar field theory described by a
(Euclidean) action in $D$ dimensional flat space and try to construct a
$D+1$ dimensional theory such that the theory is essentially controlled by the $D$
dimensional boundary theory.  The action is simply given by
\begin{eqnarray}
S = \int dt  \dot{I}(\phi) = \int dt \int d^D x  F(\phi) \dot{\phi},
\label{2.1}
\end{eqnarray}
where we have defined $F(\phi) = \frac{\delta I}{\delta \phi}$ and
the dot denotes the differentiation over $t$. Now let us perform
the canonical quantization of this theory. The momentum $p$ conjugate
to $\phi$ is given by $F(\phi)$, thereby giving rise to the primary
constraint $p - F(\phi) \approx 0$. Since the Hamiltonian is trivially
vanishing owing to the reparametrization invariance, we introduce 
the extended Hamiltonian density which is proportional to the constraint
\begin{eqnarray}
H_T = \lambda \left(p - F(\phi) \right),
\label{2.2}
\end{eqnarray}
where $\lambda$ is a Lagrange multiplier field. Then, the first-order
canonical action takes the form
\begin{eqnarray}
S = \int dt d^D x \left[ p \dot{\phi} - H_T \right] = \int dt d^D x
\left[ p \dot{\phi} - \lambda (p - F(\phi)) \right].
\label{2.3}
\end{eqnarray}
With the Poisson bracket $\{ p, \phi \}_P = 1$, we can see that the
constraint $p - F(\phi) \approx 0$ generates topological symmetry
\begin{eqnarray}
\delta \phi = \varepsilon, \ \delta p = \varepsilon 
\frac{\delta F}{\delta \phi},
\label{2.4}
\end{eqnarray}
where $\varepsilon$ is an arbitrary function. Since the constraint 
$p - F(\phi) \approx 0$ is the first-class one, using the Dirac
quantization procedure, quantum states must satisfy
\begin{eqnarray}
0 = \left(p - F(\phi)\right) \Psi = \left(- i \frac{\delta}{\delta \phi} 
- F(\phi)\right) \Psi,
\label{2.5}
\end{eqnarray}
where we have used the commutation relation $[ p, \phi] = -i$.
The unique solution to this equation is given by
\begin{eqnarray}
\Psi (\phi) = \e^{i I(\phi)},
\label{2.6}
\end{eqnarray}
up to overall normalization constant. The important knowledge gained
in this simple theory is that topological quantum field theory of
cohomological type possesses the unique quantum state described by
boundary action because of a large gauge symmetry, that is, topological 
symmetry.

Next, for later arguments, we shall consider a little more intricated
topological quantum field theory in four dimensions which was also 
discussed by Horowitz \cite{Horowitz}. The action that we consider is of form
\begin{eqnarray}
S = \int_{M_4} Tr \left( B \wedge F - \frac{1}{2} B \wedge B \right),
\label{2.7}
\end{eqnarray}
where we have used the notation of differential forms and defined
$F = dA + A \wedge A$. Here $A_\mu$ and $B_{\mu\nu}$ are a G-valued
1 form and 2 form, respectively. (For simplicity, we omit to write indices
of gauge group G.)  At first sight, it might appear that this action is
BF theory \cite{Blau}, but is in fact of cohomological type, since 
after performing the path integral over $B$ field, we obtain 
the second Chern class
\begin{eqnarray}
S = \frac{1}{2} \int_{M_4} Tr F \wedge F,
\label{2.8}
\end{eqnarray}
which is the gauge invariant action proposed by Baulieu and Singer
\cite{Baulieu} for Witten's topological quantum field theory
\cite{Witten3}.

We can express the action (\ref{2.7}) in terms of the components
\begin{eqnarray}
S = \frac{1}{4} \int_{M_4} d^4 x  \varepsilon^{\mu\nu\rho\sigma}
Tr \left( B_{\mu\nu} F_{\rho\sigma} - \frac{1}{2} B_{\mu\nu} 
B_{\rho\sigma} \right),
\label{2.9}
\end{eqnarray}
where the Levi-Civita tensor density $\varepsilon^{\mu\nu\rho\sigma}$
is defined as $\varepsilon^{0123} = +1$ and we take the metric signature
convention
$\eta_{\mu\nu} = diag(+, -, -, -)$. Since this action is linear in time
derivative it is straightforward to cast it into canonical form whose
result is given by
\begin{eqnarray}
S = \frac{1}{2} \int d^4 x  \varepsilon^{ijk}
Tr \left[ \dot{A}_i B_{jk} + A_0 D_i B_{jk} + B_{0i} 
( F_{jk}  - B_{jk}) \right],
\label{2.10}
\end{eqnarray}
where $i, j, k, \cdots$ run over spatial indices $1, 2, 3$ and 
$\varepsilon^{ijk} = +1$. Then the Hamiltonian takes the form
\begin{eqnarray}
H = - \frac{1}{2} \int d^3 x  \varepsilon^{ijk}
Tr \left[ A_0 D_i B_{jk} + B_{0i} ( F_{jk}  - B_{jk}) \right],
\label{2.11}
\end{eqnarray}
which is purely a linear combination of the constraints
\begin{eqnarray}
\varepsilon^{ijk} D_i B_{jk} \approx 0, \ F_{ij}  - B_{ij} \approx 0.
\label{2.12}
\end{eqnarray}
It is worthwhile to note that the former constraint can be derived 
from the latter ones because of the Bianchi identity 
$\varepsilon^{ijk} D_i F_{jk} = 0$, whose fact simply means that the
usual gauge symmetry generated by the former constraint is included
in topological symmetry generated by the latter constraints. Now we
are not interested in constructing the "Donaldson invariants" so
we do not take account of the "equivariant cohomology" \cite{Baulieu}
and neglect the former constraint. (Of course, even if we keep this
constraint, the result obtained below is unchanged. Incidentally, we can 
show that spatial diffeomorphisms are included in the usual gauge
transformation when the constraints are satisfied.) 

As in the previous simplest theory (\ref{2.1}), we can quantize this 
theory by using the Dirac procedure
\begin{eqnarray}
0 = \left( F_{ij}  - B_{ij} \right) \Psi = \left( F_{ij}  -  
i \varepsilon_{ijk} \frac{\delta}{\delta A_k} \right) \Psi, 
\label{2.13}
\end{eqnarray}
where we used the commutation relation $[ B_{ij}(x_0, \vec{x}),
A_k (x_0, \vec{y}) ] = i \varepsilon_{ijk} \delta (\vec{x}
- \vec{y})$. Then, the unique solution is given by the exponential
of the three dimensional Chern-Simons action
\begin{eqnarray}
\Psi = \e^{i S_{CS}(A)}.
\label{2.14}
\end{eqnarray}
Note that in comparison with the previous model (\ref{2.1}), this
result is easily expected through the famous identity
$\frac{1}{2} Tr F \wedge F = d S_{CS}(A)$. Namely, the theory under
consideration is topological quantum field theory of cohomological
type in four dimensions whose boundary theory is the three dimensional
Chern-Simons theory, so the unique quantum state is given by the
exponential of the Chern-Simons action multiplied by $i$.  
Hence, we have shown that up to overall constant the Kodama (Chern-Simons) 
state is the exact and unique quantum state of the topological 
quantum field theory (\ref{2.7}), or equivalently, (\ref{2.8}).  
This observation will be fully utilized
in the next section in order to understand why the Kodama (Chern-Simons)
state exists in Yang-Mills theory and general relativity in four
dimensions.

\section{ The Kodama (Chern-Simons) state and topological quantum field
theory }

We now consider Yang-Mills theory in four dimensional flat space-time. 
Let us start with the Yang-Mills action
\begin{eqnarray}
S = - \frac{1}{4} \int d^4 x Tr F^{\mu\nu} F_{\mu\nu}.
\label{3.1}
\end{eqnarray}
This action can be rewritten by the electric fields $E_i = F_{0i}$
and the magnetic field $B_i = \frac{1}{2} \varepsilon_{ijk} F^{jk}$
\begin{eqnarray}
S = \frac{1}{2} \int d^4 x Tr \left( {E_i}^2 - {B_i}^2 \right).
\label{3.2}
\end{eqnarray}
And up to the Gauss' law constraint the Hamiltonian takes the form
\begin{eqnarray}
H = \frac{1}{2} \int d^3 x Tr \left( {E_i}^2 + {B_i}^2 \right).
\label{3.3}
\end{eqnarray}
Given the expression for the Hamiltonian, the Schrodinger equation
is of form $H \Psi = E \Psi$ where $E$ denotes the eigenvalue of energy.
{}From the commutation relations $[ E_i(x_0, \vec{x}), A_j (x_0, \vec{y}) ] 
= - i \delta_{ij} \delta (\vec{x} - \vec{y})$, the Schrodinger equation
reduces to
\begin{eqnarray}
\frac{1}{2} \int d^3 x Tr \left(\frac{\delta}{\delta A_i} + B_i \right) 
\left(- \frac{\delta}{\delta A_i} + B_i \right) \Psi = E \Psi.
\label{3.4}
\end{eqnarray}
By using the identity $\frac{\delta S_{CS}(A)}{\delta A_i} = B^i$,
the unique quantum state with zero energy, i.e., the ground state,
is given by the Chern-Simons state \cite{Jackiw, Witten1}
\begin{eqnarray}
\Psi(A) = \e^{\pm S_{CS}(A)},
\label{3.5}
\end{eqnarray}
where $\left(\frac{\delta}{\delta A_i} \pm B_i \right) \e^{\pm S_{CS}(A)} 
 = 0$.

Now let us ask ourselves "why" this state exists in Yang-Mills theory
in four dimensions, which was one of motivations in Ref. \cite{Witten1}.
In this article, we will answer this question in a different way.
To do that, we should first notice that in order to have the Hamiltonian
with zero energy, we have to require the equations $E_i = \pm i B_i$
classically. It is then worth noting that the equations $E_i = \pm i B_i$
cause the deformation of the Yang-Mills action (\ref{3.1}) into 
a topological term,
\begin{eqnarray}
S = \pm i \int d^4 x Tr E_i B_i.
\label{3.6}
\end{eqnarray}
Given that we define the second Chern class as $c_2(A) = \frac{1}{2}
\int_{M_4} Tr F \wedge F = - \int d^4 x Tr E_i B_i$, this reduced 
action is described by $S = \mp i c_2(A)$. (The presence of the factor
$\mp i$ will be important in quantizing this theory shortly.) In other
words, we could regard the equations $E_i = \pm i B_i$ as the requirements
for picking up a topological phase among various ground states of Yang-Mills 
theory in four dimensions. 

Under the ansatze $E_i = \pm i B_i$, let us quantize the Yang-Mills
action. Since the action can be cast into the form similar to (\ref{2.7}) by
introducing the auxiliary fields $B_{\mu\nu}$
\begin{eqnarray}
S = \mp i \int_{M_4} Tr \left( B \wedge F - \frac{1}{2} B \wedge B \right),
\label{3.7}
\end{eqnarray}
using again the Dirac quantization procedure, quantum states must
satisfy
\begin{eqnarray}
0 = \left( F_{ij}  - B_{ij}\right) \Psi = \left( F_{ij}  \pm  \varepsilon_{ijk}
\frac{\delta}{\delta A_k}\right) \Psi, 
\label{3.8}
\end{eqnarray}
where we used the commutation relations $[ B_{ij}(x_0, \vec{x}),
A_k (x_0, \vec{y}) ] = \mp \varepsilon_{ijk} \delta (\vec{x}
- \vec{y})$, which can be read off from the action (\ref{3.7}). 
Here it is of importance to mention two remarks. One remark is that 
there is no appearance of $i$ in the right-handed side of the commutation 
relations owing to the existence of $\mp i$ in front of the action 
(\ref{3.7}). The other is that the commutation relations $[ B_{ij}(x_0, \vec{x}),
A_k (x_0, \vec{y}) ] = \mp \varepsilon_{ijk} \delta (\vec{x}
- \vec{y})$ are consistent with the previous ones 
$[ E_i(x_0, \vec{x}), A_j (x_0, \vec{y}) ] = - i \delta_{ij} \delta 
(\vec{x} - \vec{y})$ under the the ansatze $E_i = \pm i B_i$.
Thus, as before, we have the Chern-Simons state (\ref{3.5}) as
the unique quantum state which satisfies Eq. (\ref{3.8}). 
Of course, we can verify that, as a consistency condition, this 
state satisfies the equations 
\begin{eqnarray}
\left(E_i \mp i B_i \right) \Psi = \left(- i \frac{\delta}{\delta A_i} 
\mp i B_i \right) \Psi = 0.
\label{3.9}
\end{eqnarray}
In this way, we can understand the reason "why" the Chern-Simons 
state exists in Yang-Mills theory in four dimensions. Namely, the
reason is that with the ansatze $E_i = \pm i B_i$ the Yang-Mills theory
can be described as topological quantum field theory with the action
$S = \mp \frac{i}{2} \int_{M_4} Tr F \wedge F$ so the Chern-Simons
state corresponds to a topological state with unbroken diffeomorphism
invariance among many ground states in Yang-Mills theory.

We next turn our attention to general relativity where it is known that
there is a gravitational analog, commonly called the Kodama state 
in the community of loop quantum gravity \cite{Kodama}, of the Chern-Simons state 
in Yang-Mills theory. We begin with the chiral action with the cosmological
constant of general relativity \cite{Capovilla} \footnote{See \cite{Ikemori}
for good review of 2-form gravity.}
\begin{eqnarray}
S = \int_{M_4} \left[ R_{AB} \wedge \Sigma^{AB} - \frac{\Lambda}{6} 
\Sigma_{AB} \wedge \Sigma^{AB} - \frac{1}{2} \psi_{ABCD} 
\Sigma^{AB} \wedge \Sigma^{CD} \right],
\label{3.10}
\end{eqnarray}
where we have used the $SL(2,C)$ spinor notation for the Lorentz
group \cite{Penrose}. The indices $A, B, \cdots$, therefore, run over 
$0$ and $1$. The equations of motion derived from this action read
\begin{eqnarray}
&{}& R_{AB} - \frac{\Lambda}{3} \Sigma_{AB} - \psi_{ABCD} \Sigma^{CD} = 0,
\nn\\
&{}& D \wedge \Sigma^{AB} = 0, \nn\\
&{}& \Sigma^{(AB} \wedge \Sigma^{CD)} = 0,
\label{3.11}
\end{eqnarray}
where the round bracket denotes the total symmetrization.

It is then easy to check that the ansatze \cite{Samuel}
\begin{eqnarray}
R_{AB} = \frac{\Lambda}{3} \Sigma_{AB},
\label{3.12}
\end{eqnarray}
not only lead to a class of solutions for the equations of motion 
but also give us a solution for the quantum Ashtekar constraints
\cite{Ashtekar}. The consistency of the ansatze with the first
equations of motion in Eq. (\ref{3.11}) requires
\begin{eqnarray}
\psi_{ABCD} = 0,
\label{3.13}
\end{eqnarray}
which implies that the anti-self-dual Weyl tensor should be vanishing.
Note that in the Euclidean metric, Eq. (\ref{3.13}) is the equations
for the self-dual gravitational instantons. 

It is then interesting to see 
that under the ansatze Eq. (\ref{3.12}) and Eq. (\ref{3.13}), the action
(\ref{3.10}) becomes a topological quantum field theory of cohomological
type
\begin{eqnarray}
S = \frac{3}{2 \Lambda} \int_{M_4} R_{AB} \wedge R^{AB}.
\label{3.14}
\end{eqnarray}
Therefore, following the same line of arguments as in Yang-Mills theory
in four dimensions, it is straightforward to show that the unique quantum 
state is given by
\begin{eqnarray}
\Psi(\omega) = \e^{\frac{3}{\Lambda} S_{CS}(\omega)}.
\label{3.15}
\end{eqnarray}
where $S_{CS}(\omega)$ is the three dimensional Chern-Simons action
for spin connections $\omega$. This state is nothing but the Kodama state
for quantum gravity. Why is the factor $i$ missing in the Kadama
state?  This is traced back to the fact that the first-order
Palatini action for general relativity is actually equivalent to
$-i \int R_{AB} \wedge \Sigma^{AB}$ plus its anti-chiral part,
so precisely speaking, we should put the factor $-i$ as overall constant
in the action (\ref{3.10}) and then perform the Dirac quantization.
Consequently, we can arrive at the quantum state (\ref{3.15}).

\section{ Discussion }

In this article, we have clarified "why" the Kodama (Chern-Simons)
state exists in Yang-Mills theory and general relativity in
four dimensions. This state is associated with a topological phase
of unbroken diffeomorphism invariance and exists whenever one 
can rewrite the action into the form of the second Chern class.
In a sense, we might say that the Kodama (Chern-Simons) state
provides us a window of catching a glimpse of a relationship 
between general relativity (Yang-Mills theory) and topological 
quantum field theory.

In retrospect, it is now obvious to understand why in general relativity, 
the Kodama state is not only the WKB wave functional for the classical
solution but also an exact solution for the quantum constraints.
In the WKB approach, we evaluate the action by substituting the 
ansatze Eq. (\ref{3.12}) and Eq. (\ref{3.13}), thereby reaching 
Eq. (\ref{3.14}).  The WKB wave function can be then obtained by
exponentiating this classical action. The result of course coincides
with (\ref{3.15}), i.e., the Kodama state. As seen in the identity
$\frac{1}{2} Tr R \wedge R = d S_{CS}(\omega)$, the Kodama state
has its origin in topological quantum field theory so that this
state has a large gauge symmetry which includes both the usual
gauge symmetry and diffeomorphisms. Accordingly, the Kodama state
automatically satisfies the quantum Ashtekar constraints. 

Even if we have understood a relation between the Kodama (Chern-Simons)
state and topological quantum field theory, we have no idea whether
such a topological state is relevant to real world or not.
Of course, one of big problems in future is to clarify whether 
the Lorentzian Kodama state is normalizable under an appropriate inner 
product or not.

\begin{flushleft}
{\bf Acknowledgement}
\end{flushleft}

This work has been partially supported by Grant-in-Aid for
Scientific Research from the Japan Society for the Promotion
of Science, No. 14540277. 

\vs 1

\end{document}